\documentstyle[aps,preprint]{revtex}
\begin{document}
\draft
\author{Y. M. Vilk, Liang Chen, and A.-M. S. Tremblay}
\title{Theory of spin and charge fluctuations in the Hubbard model.}
\address{{\it D\'epartment de physique and Centre de recherche en physique du solide.}%
\\{\it Universit\'e de Sherbrooke, Sherbrooke, Qu\'ebec, Canada J1K 2R1}}
\date{\today}
\maketitle
\begin{abstract}
A self-consistent theory of both spin
and charge fluctuations in the Hubbard model is presented. It is in
quantitative agreement with Monte Carlo data at least up to intermediate
coupling $(U\sim 8t)$. It includes both short-wavelength quantum
renormalization effects, and long-wavelength thermal fluctuations which can
destroy long-range order in two dimensions. This last effect leads to a
small energy scale, as often observed in high temperature superconductors.
The theory is conserving, satisfies the Pauli principle and includes
three-particle correlations necessary to account for the incipient Mott
transition.
\end{abstract}
\hfill

The model introduced initially by Hubbard \cite{hubbard} for itinerant
magnets is now widely used for High-Temperature SuperConductors (HTSC) and
other materials with strong inter-electron interaction. Despite the apparent
simplicity of the model, its properties remain poorly understood in the
strong to intermediate coupling regimes relevant for HTSC. Much experimental
information on the magnetic fluctuations of these materials is now available
from neutron scattering and nuclear magnetic resonance. A ubiquitous feature
of the data is the presence of an unexplained small energy scale. The
one-band Hubbard model near half-filling should contain this feature if it
is the correct model for HTSC. Previous explanations\cite{chen,bulut} of the
magnetic correlations taking into account short-range quantum correlations (%
{\it T-}matrix effects) explain most of the Monte Carlo data except in the
experimentally relevant regime. No explanation of the charge structure
factor has appeared.

In this letter, we present a simple self-consistent approach to the
two-dimensional Hubbard model that gives, without adjustable parameter,
quantitative agreement with Monte Carlo data for spin and charge structure
factors and susceptibilities at {\it all} fillings up to quite strong
coupling. The approach takes into account not only the short-range quantum
effects, but also the long-range thermal fluctuations that destroy
antiferromagnetic long-range order in two-dimensions at any finite
temperature (Mermin-Wagner theorem).
This is the key physical ingredient
which leads to a small energy scale, and associated  
large correlation length, in the magnetic fluctuations. 
Previous
approaches which included the long-range thermal fluctuations in the 2D
Hubbard model were never applied to the incommensurate case relevant for
HTSC. Furthermore, they are based on mode-coupling theory\cite{moriya},
which neglects charge fluctuations and does not include the effect of
short-range quantum correlations, which are important not only for a
quantitative description of the model but also for determining the nature of
the ground state.

We first present our approach, discuss physical consequences, and finally
compare with Monte Carlo data. We consider the one-band Hubbard model with
on-site repulsive $U$. Our approach is motivated by the Local Field
Approximation (LFA), which was successful in the electron gas\cite{singwi}.
We start from the equation of motion for the particle-hole operator $\rho
_\sigma \left( \vec l,\vec l^{\prime }\right) =c_{\vec l,\sigma }^{\dagger
}c_{\vec l^{\prime },\sigma }$ in a weak external field $\phi _{\vec l%
,\sigma }$ which is coupled to the partial density operator $n_{\vec l%
,\sigma }=\rho _\sigma \left( \vec l,\vec l\right) =c_{\vec l,\sigma
}^{\dagger }c_{\vec l,\sigma }$ \cite{pines}. After simple transformations,
the term that contains the interaction in the equation of motion for $\rho
_\sigma \left( \vec l,\vec l^{\prime }\right) $ is of the form,

\begin{equation}
\label{correlator}U\quad [\langle \rho _\sigma \left( \vec l,\vec l^{\prime
}\right) n_{\vec l,-\sigma }\rangle -\langle \rho _\sigma \left( \vec l,\vec 
l^{\prime }\right) n_{\vec l^{\prime },-\sigma }\rangle ]. 
\end{equation}
All operators contain the same time label, which is not explicitly written.
The usual Random Phase Approximation (RPA) corresponds to the neglect of
two-particle correlations, namely one approximates $\langle \rho _\sigma
\left( \vec l,\vec l^{\prime }\right) \rho _{-\sigma }\left( \vec l,\vec l%
\right) \rangle $ by $\langle \rho _\sigma \left( \vec l,\vec l^{\prime
}\right) \rangle \langle \rho _{-\sigma }\left( \vec l,\vec l\right) \rangle 
$. This is clearly a poor approximation for on-site interactions because, as
can be seen from (\ref{correlator}), three of the four creation or
annihilation operators in the correlator $\langle \rho _\sigma \left( \vec l,%
\vec l^{\prime }\right) \rho _{-\sigma }\left( \vec l,\vec l\right) \rangle $
are on the same lattice site $\vec l$. There is thus a strong correlation
between two particle-hole pairs even when $|\vec l-\vec l^{\prime }|>>1$ 
(the lattice constant is taken to be unity). 
We make use of this specific feature of the
on-site interaction and neglect the dependence of the correlation
coefficient on the lattice index $l^{\prime }$ which appears only once in
the two-particle correlator. Mathematically, our {\it ansatz} is, 
\begin{equation}
\label{decoupling}\langle \rho _\sigma \left( \vec l,\vec l^{\prime }\right)
n_{\vec l,-\sigma }\rangle =g_{\uparrow \downarrow }\left( \vec l,\vec l%
\right) \langle \rho _\sigma \left( \vec l,\vec l^{\prime }\right) \rangle
\langle n_{\vec l,-\sigma }\rangle 
\end{equation}
where 
\begin{equation}
\label{g}
g_{\sigma \sigma ^{\prime }}\left( \vec l,\vec l^{\prime }\right) =
\frac{[\langle n_{\vec l%
,\sigma }n_{\vec l^{\prime },\sigma ^{\prime }}\rangle -\langle n_{\vec l%
,\sigma }\rangle \delta _{\sigma ,\sigma ^{\prime }}\delta _{\vec l,\vec l%
^{\prime }}]}{(\langle n_{\vec l,\sigma }\rangle \langle n_{\vec l^{\prime
},\sigma ^{\prime }}\rangle )}
\end{equation}
 is the pair correlation function 
between electrons of spin $\sigma $ and $%
\sigma ^{\prime }$ on the respective lattice sites $\vec l$ and $\vec l%
^{\prime }$.

It is important to realize that the pair correlation function in (\ref
{correlator}) and (\ref{decoupling}) cannot yet be taken equal to its
equilibrium value $g_{\uparrow \downarrow }\left( \vec l,\vec l;t\right)
\neq g_{\uparrow \downarrow }\left( 0\right) $ because of the weak external
field $\phi _{\vec l,\sigma }$. In linear response the most general form for 
$g_{\uparrow \downarrow }\left( \vec l,\vec l\right) \quad $is:

$$
g_{\uparrow \downarrow }\left( \vec l,\vec l;t\right) =g_{\uparrow
\downarrow }\left( 0\right) +\sum_{l^{\prime }}\int dt^{\prime }\frac{\delta
g_{\uparrow \downarrow }\left( \vec l,\vec l;t\right) }{\delta \langle n_{%
\vec l^{\prime },\uparrow }\left( t^{\prime }\right) \rangle }\delta \langle
n_{\vec l^{\prime }}\left( t^{\prime }\right) \rangle 
$$
where $n_{\vec l}\left( t\right) =n_{\vec l,\uparrow }\left(
t\right) +n_{\vec l,\downarrow }\left( t \right) $ and 
we use that in the paramagnetic state $\frac{\delta
g_{\uparrow \downarrow }\left( \vec l,\vec l;t\right) }{\delta \langle n_{%
\vec l^{\prime },\uparrow }\left( t^{\prime }\right) \rangle }=\frac{\delta
g_{\uparrow \downarrow }\left( \vec l,\vec l;t\right) }{\delta \langle n_{%
\vec l^{\prime },\downarrow }\left( t^{\prime }\right) \rangle }\equiv
g_{_{\uparrow \downarrow }}^{\prime }\left( \vec l-\vec l^{\prime
},t-t^{\prime }\right) $. Note that the terms describing the response of $%
g_{\uparrow \downarrow }\left( \vec l,\vec l;t\right) $ to the external
field enter the equation of motion only in a form that is symmetric in spin
indices. Since the $z-$component of spin $S_z=n_{\uparrow }-n_{\downarrow 
\text{ }}$is antisymmetric in $n_\sigma $ and all equations are linear, it
immediately follows that $g_{_{\uparrow \downarrow }}^{\prime }\left( \vec l-%
\vec l^{\prime },t-t^{\prime }\right) $ enters the equation for charge but
not for spin. This important simplification occurs because the Pauli
principle precludes the appearance of terms like $\frac{\delta
g_{\uparrow \uparrow }}{\delta \langle n_{\uparrow }\rangle }\neq \frac{%
\delta g_{\uparrow \uparrow }}{\delta \langle n_{\downarrow }\rangle }$ in
the case of on-site interaction. After standard transformations (see for
example \cite{pines}) the spin and charge susceptibilities have the RPA form
but with a renormalized effective interaction, which is different for spin $%
(U_{sp})$ and charge $(U_{ch})$: 
\begin{equation}
\label{Uren}
U_{sp}=g_{\uparrow \downarrow }\left( 0\right) U;\; U_{ch}=\left(
g_{\uparrow \downarrow }\left( 0\right) +\delta g_{\uparrow \downarrow
}\left( \omega ,\vec q\right) \right) U 
\end{equation}
where $\delta g_{\uparrow \downarrow }\left( \omega ,\vec q\right)
=g_{\uparrow \downarrow }^{\prime }\left( \omega ,\vec q\right) \frac n2$, $%
g_{\uparrow \downarrow }^{\prime }\left( \omega ,\vec q\right) $ is Fourier
transform of $g_{_{\uparrow \downarrow }}^{\prime }\left( \vec l-\vec l%
^{\prime },t-t^{\prime }\right) $ and $n$ is the band filling (half-filled
case corresponds to $n=1$). $\delta g_{\uparrow \downarrow }\left( \omega ,%
\vec q\right) $ is a three-particle correlation function, so that further
simplification will be needed to calculate the charge susceptibility $\chi
_{ch}\left( \omega ,\vec q\right) $. However, no further approximation is
needed for the spin susceptibility $\chi _{sp}\left( \omega ,\vec q\right) $%
! Indeed, due to the Pauli principle, $g_{\uparrow \uparrow }\left( 0\right)
=0$ so the spin part of the problem may be closed by using $g_{\uparrow
\downarrow }\left( 0\right) =-2g_{sp}(0)$ with $g_{sp}(l,l^{\prime })\equiv
(g_{\uparrow \uparrow }(l,l^{\prime })-g_{\uparrow \downarrow }(l,l^{\prime
}))/2$ and the Fluctuation-Dissipation Theorem (FDT) for spin:

\begin{equation}
\label{spin}
\begin{array}{c}
g_{sp}\left( l,l^{\prime }\right) = 
\frac 1n\int \frac{d^2q}{\left( 2\pi \right) ^2}\left[ S_{sp}\left( \vec q%
\right) -1\right] e^{i\vec q\cdot (l-l^{\prime })} \\ S_{sp}\left( \vec q%
\right) =\frac Tn\;\sum_{i\omega _m}\frac{\chi _0\left( i\omega _m,\vec q%
\right) }{1-(U_{sp}/2)\chi _0\left( i\omega _m,\vec q\right) } 
\end{array}
\end{equation}
The first equation is the definition of the spin structure factor; the
second is a convenient form of the FDT with temperature $T$ and bosonic
Matsubara frequencies $i\omega _m$. The integral is over the first
Brillouin zone. The definition of $\chi _0\left( i\omega _m,\vec q\right)$ is
the same as in Ref. \cite{pines}. 

For $\delta g_{\uparrow \downarrow }\left( \omega ,\vec q\right) $, we use
the simplest {\it ansatz, }namely, that it is a constant $\delta g_{\uparrow
\downarrow }$, which we determine self-consistently using the Pauli
principle $g_{\uparrow \uparrow }\left( 0\right) =0$, the definition for the
static charge structure factor 
\begin{equation}
\label{charge}g_{ch}\left( l,l^{\prime }\right) =1+\frac 1n\int \frac{d^2q}{%
\left( 2\pi \right) ^2}\left[ S_{ch}\left( \vec q\right) -1\right] e^{i\vec q%
\cdot (l-l^{\prime })} 
\end{equation}
and the FDT for charge. We thus have a simple theory with only two
parameters $g_{\uparrow \downarrow }\left( 0\right) $ and $\delta
g_{\uparrow \downarrow }$ that are found self-consistently. 
It can be explicitly checked
that charge, spin, and energy are conserved. 
As with all self-consistent theories, the
usefulness of the approach can be judged only {\it a posteriori }by
comparison with numerical or exact results. We provide such comparisons
later in the paper.\\
\indent 
The absence of a magnetic phase transition at any finite temperature in 2D
follows immediately from the above approach. Define the mean-field critical
value of $U$ by $U_{mf,c}=2/\chi _0\left( 0,\vec q_{\max }\right) $ where $%
\chi _0\left( i\omega _m,\vec q_{\max }\right) $ is the susceptibility of
non-interacting electrons and $\vec q_{\max }$ is the value of $\vec q$ at
which the static susceptibility has its maximum. If $U_{sp}=g_{\uparrow
\downarrow }(0)U$ in (\ref{spin}) was large enough for the transition to
occur, namely $\delta U=[U_{mf,c}-g_{\uparrow \downarrow }\left( 0\right)
\,U]=0$, then the $\vec q-$integral for the static susceptibility ($\omega
_m=0$) in the expression for $g_{sp}(0)$ (\ref{spin}) would diverge
logarithmically so that $g_{\uparrow \downarrow }\left( 0\right)
=-2g_{sp}(0) $ would become negative, in obvious contradiction with $\delta
U=0$. Hence, in our approach, magnetic fluctuations always push the value of 
$g_{\uparrow \downarrow }\left( 0\right) $ away from its critical value $%
g_{\uparrow \downarrow }^{\left( c\right) }\left( 0\right) =U_{mf,c}/U$ at
any finite temperature. Furthermore, for a wide range of values of $%
U>U_{mf,c}$ and $T<T_{mf,c}$ , the system will be quite close to magnetic
instability $(\delta U\sim 0)$, providing the basis for a generic
explanation of the small energy scale observed in HTSC. In the regime
in which the temperature is larger than this small energy scale, the
correlation length grows exponentially $\xi \propto \tilde \xi \propto
e^{const/T}$ \thinspace $(\tilde \xi ^{-2}\equiv \delta U/U_{mf,c}$),
reflecting the logarithmic divergence mentioned above. This is typical
behavior for systems at their lower critical dimension. When a real
quasi-two-dimensional system enters this regime, small three-dimensional
effects can easily stabilize a long-range order.

We digress briefly to speculate on how the phase diagram of a system with
weak three-dimensional effects would then look. We neglect effects, such as
disorder, which may become important when the small energy scale $\delta U$
appears. We define a quasi-critical temperature $T_{qc}$ as the temperature
at which the enhancement of the magnetic susceptibility $\tilde \xi ^2=\chi
_{sp}\left( 0,\vec q_{\max }\right) /\chi _0\left( 0,\vec q_{\max }\right) $
is of order $500$. Calculations for the nearest-neighbor Hubbard model show
that $\tilde \xi ^2$ increases by an order of magnitude when the temperature
is reduced below $T_{qc}$ by as little as $T_{qc}-T\sim 0.01$ (all energies
are in units of the hopping integral $t$). In our theory, the emerging
long-range order is determined by the position of the maximum of $\chi
_0\left( 0,\vec q\right) $. At $T=0$, as soon as $n\neq 1$, two-dimensional
Fermi-surface effects \cite{chen2} lead to a maximum with a cusp-like
singularity in the $(\pi ,\pi )-(\pi ,0)$ direction. The situation is
different at finite temperature where the maximum can be at $\vec q_{\max
}=(\pi ,\pi )$ even for $n<1$. Fig.~1 shows a rough magnetic phase diagram
obtained by approximating $T_c$ by $T_{qc}$ (with $U=2.5$). The inserts show
the dependence of $q_{y,\max }$ ($\vec q_{\max }=(\pi ,q_{y,\max })$ ) and
of the enhancement factor $\tilde \xi ^2$ on temperature for three different
fillings. The filling $n=0.93$ is marginal in the sense that the shift of $%
\vec q_{\max }$ from $(\pi ,\pi )$ occurs when the enhancement $\tilde \xi
^2 $ is already quite large. We would expect then that long-range order
would be antiferromagnetic (AFM) for $n>0.93$ and incommensurate
spin-density wave for $0.89<n<0.93$.\\
\indent Let us now discuss the limit $T\rightarrow 0$. In Fig. 2, we show the
renormalized spin interaction $U_{sp}$ as function of the filling $n$ for $%
U=2.5$ and $U\rightarrow \infty $, together with $U_{mf,c}\left( n\right) $.
Remarkably, at low filling the value of $U_{sp}$ saturates to $%
U_{sp}\approx 3.2$ as the bare interaction increases $U\rightarrow \infty $.
This quantum effect was anticipated by Kanamori\cite{kanamori}, \cite{chen}
who argued that the largest value of $U_{sp}$ is proportional to the kinetic
energy cost to put a node in the two-body wave-function where two electrons
overlap. At a sufficiently large filling, when $U_{sp}(n)$ starts to follow $%
U_{mf,c}\left( n\right) $ for $T\rightarrow 0$, the paramagnetic state has
an instability at exactly $T=0$. The existence of an upper limit for $U_{sp}$
leads to the existence of a lower limit for the filling $n_{c,\min }\approx
0.685$ below which there is no magnetic phase transition at any $U$. This,
in turn, means that only spin-density waves with $\vec q=(\pi ,q_y),\quad
q_y/\pi \subset [0.74,1]$ are possible. In particular, the ferromagnetic
state does not exist in the Hubbard model on a square lattice\cite{chen} and
there is also a temperature $T \approx 0.5$ above which there is no
exponential regime for $\xi (T)$ at any $U$ where our theory applies.\\
\indent In the rest of this letter, we compare our theoretical results for infinite
lattice with Monte Carlo simulations. Finite-size effects are expected to
become important in simulations when the correlation length becomes
comparable with system size . The largest size effects are thus expected in $%
S_{sp}\left( \pi ,\pi \right) $ at half-filling ($n=1$). This is shown in
Fig. 3. The Monte Carlo data\cite{white} follow our theoretical curve (solid
line) until they saturate to a size-dependent value. We checked that
finite-size effects for $S_{sp}\left( \vec q\right) $ away from the
antiferromagnetic wave vector $\vec q=\left( \pi ,\pi \right) $ are much
smaller, so that, even for $8\times 8$ systems, theory and simulations agree
very well for all other values of $\vec q$ (not shown). Finite-size effects
in the spin structure factor $S_{sp}\left( \vec q\right) $ are not too
important away from half-filling for the parameters shown on Fig. 4a-d.
Obviously though, finite-size simulations cannot reproduce the small
incommensurability captured by our theory at $n=0.8$ on Fig.~4d.
 
Figures. 4b and 4c show that, even for relatively strong coupling $(U\sim 8)$%
, the theory agrees very well with {\it both} spin $S_{sp}\left( \vec q%
\right) $ and charge $S_{ch}\left( \vec q\right) $ structure factors.
However the theory should eventually break down for $U\rightarrow \infty $.
This can be seen in the half-filled case from the fact that, for $%
U\rightarrow \infty $, the antiferromagnetic susceptibility remains constant
in our theory while mapping to the Heisenberg model with $J=4t^2/U$ shows
that it should decrease with $U$. It seems, however, that large-$U$
asymptotic is reached for values of $U$ much larger than the bandwidth.\\
\indent Fig. 4a shows that our theory reproduces the important qualitative fact that
the charge structure factor $S_{ch}\left( \vec q;n\right) $ depends on
filling in a non-monotonous manner. The decrease of $S_{ch}\left( \vec q%
\right) $ towards half-filling is a signature of the incipient Mott
transition. The effect can be seen because our approach takes into account
both three-particle correlations and the Pauli principle. Writing the Pauli
principle as a sum-rule $\Sigma _{\vec q}[S_{ch}\left( \vec q\right)
+S_{sp}\left( \vec q\right) ]=2\,\Sigma _{\vec q}\,S_0\left( \vec q\right) $,
the parameter $\delta g_{\uparrow \downarrow }$, which partially takes
into account three-particle correlations, must increase close to
half-filling in order to reduce $S_{ch}\left( \vec q\right) $ and compensate
for the increase in the contribution of the spin structure factor.

Our theory also explains the good fit of the dynamical spin susceptibility $%
\chi _{sp}\left( i\omega _m,\vec q\right) $ obtained by Bulut {\it et al.} 
\cite{bulut} using RPA with $U_{ren}=2$. 
Indeed, for $U=4$, $n\approx 0.87$ on $8\times 8$ clusters,
our calculations give $U_{sp}=2.05$ with very little dependence on
temperature. Bulut et al. have also shown that their Monte Carlo 
data for the self-energy $\Sigma
\left( i\omega _m,\vec q\right) $ can be reasonably well fitted by the
Berk-Schrieffer formula\cite{berk} with the same $U_{ren}=2$. In our
approach, the expression for $\Sigma \left( i\omega _m,\vec q\right) $ in
terms of susceptibilities should come at the next level of approximation.
Bulut {\it et al.} \cite{bulut2} have also fitted a number of experiments in
HTSC by fine-tuning the value of $U_{ren}$ close to a magnetic instability $%
(\delta U\sim 0)$. In our approach, a wide range of bare values of $U$
naturally renormalizes to such a situation.

In conclusion, imposing the Pauli principle as well as self-consistency
through the fluctuation-dissipation theorem, we have formulated a simple
theory that also satisfies conservation laws and gives, without adjustable
parameter, a quantitative explanation of Monte Carlo data for both spin and
charge structure factors as well as susceptibilities up to intermediate
coupling. Both short-wavelength quantum renormalization effects and
long-wavelength thermal fluctuation effects, which destroy long-range order
in two-dimensions, are accounted for. The latter effect naturally leads to a
small energy scale for a wide range of parameters, possibly giving a
microscopic origin for the small energy scale observed in experiments on
high-temperature superconductors.

We thank A.E. Ruckenstein for useful discussions. This work was partially
supported by the Natural Sciences and Engineering Research Council of Canada (NSERC),
the Fonds pour la formation de chercheurs et l'aide \`a la recherche from
the Government of Qu\'ebec (FCAR), and (A.-M.S.T.) the Canadian Institute of
Advanced Research (CIAR).

\figure{ Fig. 1. Approximate phase diagram of the quasi-2D Hubbard model 
with $T_c$ approximated by $T_{qc}$.  The insert  shows the
 temperature dependence 
of $q_{max}(T)$ and the enhancement
 factor $\tilde{\xi}^2=\chi_{sp}(0,q_{max})/\chi_0(0,q_{max})$
for three different fillings $n$. \label{phd}} 
\figure{Fig. 2. Filling dependence of $U_{sp}$ and $U_{mf,c}$ 
as $T \rightarrow 0$. \label{usp}} 
\figure{ Fig. 3. Temperature dependence of $S_{sp}(\pi,\pi)$ at half-filling $n=1$.
The solid line is our theory; symbols are Monte Carlo data from 
Ref. \cite{white}. \label{sspn1t}} 
\figure{Fig. 4.  Wave vector ($\vec{q}$) dependence of the 
spin and charge structure factors 
for different sets of parameters. Solid lines are our theory; symbols 
are our Monte Carlo data. Monte Carlo data for $n=1$ and $U=8$ are
 for $6 \times 6$ clusters and $T=0.5$; all other data are for $8 \times 8$ clusters
and $T=0.2$. Error bars are shown only when significant. \label{spch}}
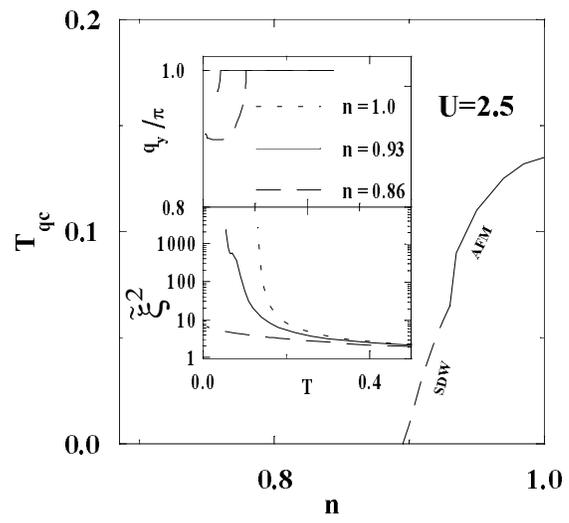
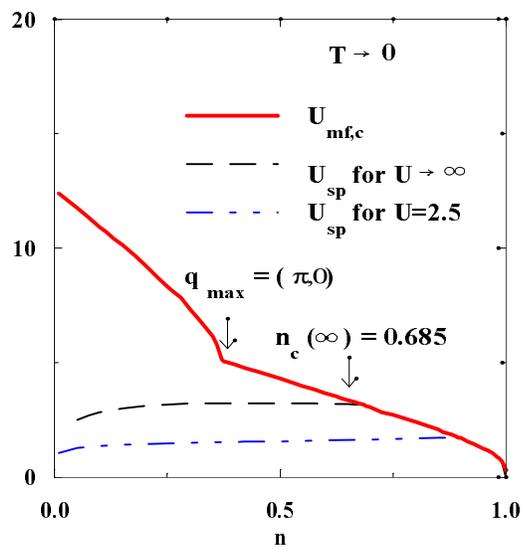
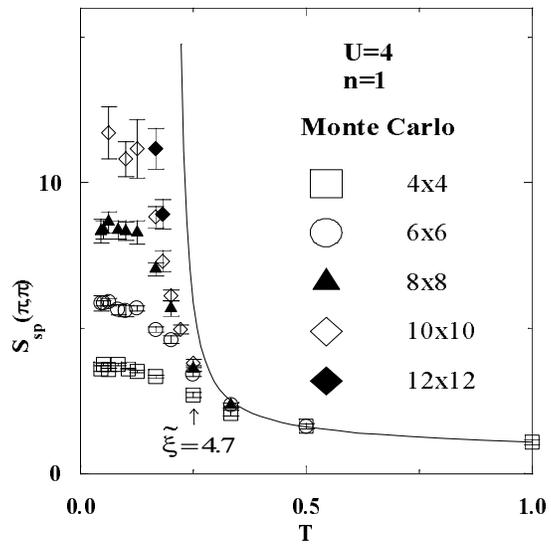
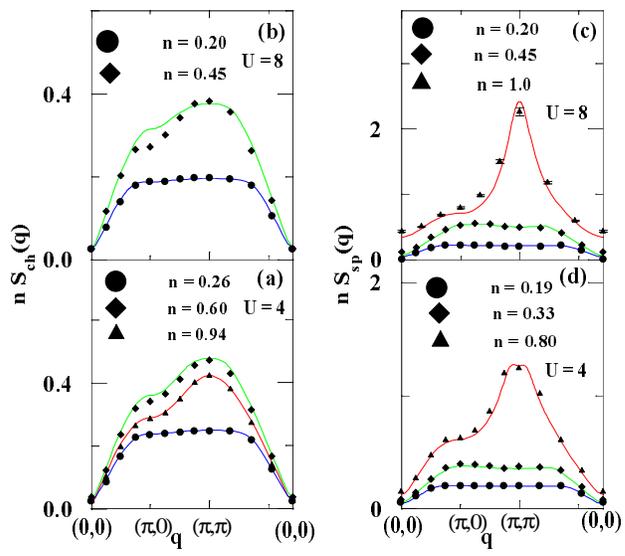
\end{document}